# Providing a hybrid cryptography algorithm for lightweight authentication protocol in RFID with urban traffic usage case

V. Chegeni [1], H. Haj Seyyed javadi [2*], M.R Moazami Goudarzi [3], A. Rezakhani [4]

[1] Department of Computer Engineering, Borujerd Branch, Islamic Azad University, Borujerd, Iran
[2] Department of Mathematics and Computer Science, Shahed University, Tehran 3319118651, Iran
[3] Department of Mathematics, Borujerd Branch, Islamic Azad University, Borujerd, Iran
[4] Department of Computer Engineering, University of Ayatollah Borujerdi, Borujerd, Iran
* h.s.javadi@shahed.ac.ir

**Abstract:** Today, the Internet of Things (IoT) is one of the emerging technologies that enable the connection and transfer of information through communication networks. The main idea of the IoT is the widespread presence of objects such as mobile devices, sensors, and RFID. With the increase in traffic volume in urban areas, the existing intelligent urban traffic management system based on IoT can be vital. Therefore, this paper focused on security in urban traffic based on using RFID. In our scheme, RFID tags chose as the purpose of this article. We, in this paper, present a mutual authentication protocol that leads to privacy based on hybrid cryptography. Also, an authentication process with RFID tags is proposed that can be read at high speed. The protocol has attempted to reduce the complexity of computing. At the same time, the proposed method can withstand attacks such as spoofing of tag and reader, tag tracking, and replay attack.

## 1. Introduction

Nowadays, there is a need for automatic identification and data collection of elements without the need for human intervention to enter information in many industrial, scientific, service, and social fields. IoT technology is one of the areas under study that is a fast-growing pattern. The main idea of the IoT is the widespread presence of objects such as mobile devices, sensors, and RFID, etc. in a network that is assigned a unique IP address for each of these devices [1]. The Internet of Things (IoT) is defined as a network of smart devices that share information by interacting with one another.

IoT is gradually becoming a significant part of different aspects of our lives. It is used in smart homes, wearable devices, healthcare, etc. Its wide range of applications yields common data, such as the significant value of users' private information. Hence, the security of this information is very consequential. To provide security for the IoT, several factors, such as data confidentiality, data integrity, authentication, access control, privacy, etc. are required [2].

The IoT is a framework for the smart city and smart energy management systems that are used widely today. The growing population and increased vehicles lead to the main challenges in urban life. One of the crucial issues in smart cities is the use of intelligent urban traffic management. One of the significant challenges in IoT is security. Therefore, this paper focused on security in urban traffic based on IoT. In our scheme, RFID tags chose as the purpose of this article. An essential part of RFID in IoT focus on security, especially in IoT networks [3]. Security of the IoT can be summarized in four cases [4]:

1. **Attack Resistance:** The network must be resistant to the disruption of a node and be able to adapt to attack conditions.
2. **Data authentication:** The received data and addresses must be authenticated to prevent malicious data from being entered into the network and devices.
3. **Access control:** Access to data must be controlled so attackers cannot access it without permission.
4. **Data privacy:** Keep in mind that only network administrators can see the names and information of users, and others cannot easily access this information.

In this study, we used RFID technology to consider the application of the proposed authentication protocol. RFID technology has received much attention due to its widespread use in the modern era, so we also investigated the issue and applied our protocol to technology and considered the use of traffic management. RFID technology is the concept of radio wave identification. In this technology, an object is identified by sending its identifier via radio waves. These systems consist of three main components: tag, reader, and central server [1] [5].

The sensors on the IoT, in addition to the senses, can process and store sensed events. They can even intelligently recognize if a sensed event is a repeating one. IoT combines multiple technologies such as RFID, wireless sensor networks (WSN), NFC, etc. IoT is responsible for data processing, manipulation, and decision-making. IoT sends the data to the internet in only one hop. First, in IoT, routing is not implemented. Sensors send their data directly to the internet because they have an internet connection. In IoT, each device is identifiable with a unique ID, that is, its IP address [6].

We, in this paper, use a mutual authentication protocol based on hybrid cryptography (the combination of symmetric and asymmetric cryptography algorithms) with RFID tags to improve reading speed in RFID. In our scheme, Advanced Encryption Standard (AES) algorithm was used to encrypt the data block; then, Elliptic-curve cryptography (ECC) algorithm was used to encrypt the AES secret key, which is sent between the tag and the central server. In our approach, the protocol has attempted to increase the speed of computing on a central server. We examine the security of this protocol against several possible attacks on the use of urban traffic. The proposed approach has been analyzed and compared both analytically and experimentally with other cryptographic algorithms. Our approach illustrates that increased to detect a high percentage of RFID tags in the environment.

The rest of the paper is organized as follows. In section 2, related work is described. The components of the RFID system are explained in section 3. In the fourth section, the proposed authentication protocol is described, and in section 5 and 6, we introduce and analyses the proposed method evaluation and summarization, respectively. Finally, conclusion of this paper is stated in Section 7.

## 2. Related Work

Mutual authentication between IoT devices and IoT servers is an essential part of secure IoT systems. Password-based authentication mechanisms, which are widely used, are vulnerable to side attacks and vocabulary. Shah et al. [7], proposed a cross-authentication mechanism based on multiple keys (or multiple passwords). In the approach proposed in that article, there is a shared password vault between the IoT server and the IoT device, which is a set of corresponding keys. In this regard, after each use of the keys in each session, the keys are removed. Then, new keys are replaced, which ensures security in the authentication.

Data generated from various smart devices in the IoT environment is one of the biggest concerns. Cloud computing has emerged as a critical technology for processing such a large database repository of various types of devices in the IoT environment. However, IoT devices store private information on distributed private cloud servers so that only authenticated users can access sensitive information from the cloud server. Amin et al. [8], focus on all these points, the security vulnerabilities of the multicast cloud server environment are first outlined. Then, an architecture for these challenges is proposed that applies to distributed cloud environments, whereby a smart card authentication protocol is provided in which the registered user can fully access all cloud servers. Private access to all private information and access is secure.

In the smart grid, bi-directional communication is essential. Also, security and privacy are the essential requirements that should be provided in the communication. Due to the intricate design of smart grid systems, adversaries can attack the smart grid system, which can lead to deadly problems for customers. Mahmood et al. [9] proposed a lightweight message authentication scheme for smart grid communications and claimed that it satisfies the security requirements. However, Bayat et al. [10] found that Mahmood et al.'s scheme has some security vulnerabilities, and it has not adequate security features to be utilized in the smart grid. To address these issues, [10] propose an efficient and secure lightweight privacy-preserving authentication scheme for a smart grid.

Alimohammadi et al. [11] proposed a data-sharing scheme for cloud storage. The data owner encrypts the files to upload on the cloud. Each user should send a request to the owner and get the authorization to download the files. The entities authenticate each other in such communications. The user introduces itself to the cloud via the authorization. If the cloud confirms the authorization, it allows the user to download the requested files. In this article, vulnerability against DoS and impersonation attacks are investigated. The results demonstrate that the proposed approach is suitable for data sharing in dynamic cloud storage.

A group-based secure, lightweight handover authentication (GSLHA) protocol was proposed for M2M communication in LTE and future 5G networks [12]. A group of MTC devices (MTCDs) and a new eNodeB (eNB) when simultaneously enter the coverage of the eNB mutually authenticates in the proposed protocol. The results indicate that the proposed protocol has been able to achieve all the security goals and overcome various attacks.

Three methods are used by a KPS that are random, deterministic, and hybrid ones. The first schemes require that the keys be selected randomly from a key pool and stored in each object. This method will not guarantee the direct communication of each two nodes. Lack of direct communication leads to the creation of a path between the two nodes, which reduces the speed of communications. Also, to design a key pool and key rings to achieve better key connectivity, a deterministic method should be used. A combination of both the deterministic and random approaches creates a hybrid method that can be used to improve scalability and resiliency [13].

The term security is a vital issue in any sensor network. Therefore, In the sensor networks, key management is considered the main security service. Due to the limitations on sensor nodes, traditional key management and key agreement techniques do not fit with sensor networks. A new key pre-distribution scheme was proposed by Anzani [14] using multivariate polynomials to establish the pairwise key in sensor networks. Based on this approach, the combinatorial design theory must be applied in the multivariate key pre-distribution scheme. In this scheme, sensor nodes store the common multivariate polynomials before deploying the network using the identifier of sensors and the combinatorial design.

Sklavos et al. [15] proposed, the current cryptographic methods such as the Advanced Encryption Standard (AES) and the Elliptic Curve Cryptography (ECC) are expounded, and their functionality with their advantages and disadvantages are discussed together. Also, this paper highlights the need for more flexible cryptographic suites.

Authentication is an essential security requirement for Session Initiation Protocol (SIP). An improved authentication scheme concerning SIP is proposed, in which they claimed that their scheme is secure against various security attacks [16]. However, Arshad et al. [17], conclude that the scheme in [16] is vulnerable to user impersonation attacks. Furthermore, they proposed a new authentication and key agreement scheme for SIP using the Elliptic Curve Cryptography (ECC). Their scheme includes four phases: system setup phase, registration phase, password change phase, and authentication and key agreement phase. Security and performance analyses illustrate that the proposed scheme is secure against security attacks of various types and has low computation cost compared to previously proposed schemes.

The 5G network is a key factor in meeting the ever-increasing demand for IoT services, including high-speed data, connecting multiple devices, and delaying service. To meet these demands, fog computing has been envisioned as a promising solution in 5G service-oriented architecture to enhance security. However, the security paradigms that enable the authentication and confidentiality of 5G communications for IoT services to remain inevitable. Ni et al. [18] proposed an efficient and secure service-oriented authentication framework with support for fog computing for IoT services on 5G networks. Correctly, users can connect to 5G networks and remain completely anonymous using the proposed method while also accessing IoT services. Also, session keys are negotiated among users, fog, local guests, and IoT servers to provide secure access to service data in fog cache and low-latency remote servers.

## 3. RFID System Components

An RFID system consists of three main components: tag, reader, and a central server, to identify elements. The reader sends a signal to the RFID tag, requesting its information, and the RFID tag responds to the signal by sending the information to the reader. The reader then sends this information to the central server to identify the RFID tag.

A radio tag consists of an antenna for receiving and transmitting signals, and an integrated circuit for signal conversion as well as information storage and processing. Each tag contains the unique identifier of the element to which it is attached. This ID is sent in response to a reader request to identify the element containing it [19] [20]. RFID tags divide into three general categories of active, passive, and semi-passive tags. The following will describe the features of each.

### 3.1. Active RFID tags

These tags contain a built-in battery that allows them to perform more complex tasks. These tags can perform internal calculations in the absence of a reader. These tags even initiate the relationship between the tag and the reader [21].

### 3.2. Passive RFID tags

Another type of tag is the passive tag that does not have an internal battery. These tags derive their energy needed to perform all the calculations from the reader. Therefore, these tags can never initiate the relationship between themselves and the reader [21].

### 3.3. Semi-passive RFID tags

The third type of tags, which are called semi-passive, are similar to active tags containing built-in batteries, except that they use energy only for internal calculations [22].

## 4. The Proposed Authentication Protocol

In this study, we used an improved version of the Q-Protocol anti-collision algorithm. The Q-protocol is a Dynamic Framed Slotted ALOHA (DFSA)-type protocol that modifies frame size using feedback from the last frame accomplished. The ALOHA algorithm has many advanced versions, for example, Slotted ALOHA, Framed Slotted ALOHA, and Dynamic Framed Slotted ALOHA (DFSA). ALOHA algorithm is a simple anti-collision method based on TDMA. When the tag reaches the interrogation area of a reader, the tag will transmit the data immediately, and when more than one tag response at the same time, the collision occurs. DFSA algorithm changes the frame size for tag identification. The frame size is determined using the information such as the number of slots used to identify the tag and the number of the slots collided. For example, when the number of the slots collided is larger than the upper limit, the reader will add the number of slots in one frame; when the number of the empty slots is smaller than the lower limit, the reader will decrease the number of slots in one frame [23] [24] [25].

According to the DFSA protocol, the reader sends a 22-bit Query command to the tag. The Query command contains one or more frames, each frame consisting of some slots. The reader determines the value of parameter $Q$ by sending this command to specify the frame size ($2^Q - 1$).

According to which the tags select a slot between zero and frame size $2^Q - 1$, the tag sends the output 16-bit $RN16$ value to the channel reader for the channel reservation. If a tag successfully transmits its RN16 without error or collision, the reader sends 18-bit ACK to the tag. After receiving the ACK from the reader, the tag sends the data to the reader.

Due to the randomness of the slot selection process, one of the following three situations may occur for each slot:
1. None tag did not have a selected slot.
2. Two or more tags have selected a slot.
3. Only one tag has selected a slot.

In the first case, the slot is idle. In the second case, a collision occurs, and these tags again participate in the identification process in the next frame. In the third case, the slot is assigned to the tag in which the slot sends its data to the reader. This method will work best if the number of slots is approximately equal to the number of tags in the environment. Because in this case, the slot does not remain idle, and also no collision occurs.

### 4.1. Hybrid Cryptographic Algorithm

In [26], two asymmetric cryptographic algorithms, such as ECC [27] [28] and Rabin [29], are considered to perform the authentication. Asymmetric cryptographic algorithms are a powerful tool for maintaining system security. However, the use of these algorithms in RFID with limited resources is a challenge. These algorithms have cumbersome and time-consuming computations that make them difficult to use on RFID. Hence, lighter algorithms are designed to be implemented in such systems.

Our scheme presents the authentication protocol base on a combination of symmetric and asymmetric cryptography algorithms. Furthermore, the advantages of both cryptographic algorithms are used. This way, the reader receives the information encrypted by the tag and sends it to the central server. The central server decrypts this message and authenticates the tag. It then releases the information encrypted to be written on the tag and sends it to the reader, and the reader sends it to the tag. The protocol is a mutual authentication protocol in which the tag and the central server authenticate each other, and the reader is the only link between them.

In our scheme, Advanced Encryption Standard (AES) algorithm was used to encrypt the data block; then, Elliptic-curve cryptography (ECC) algorithm was used to encrypt the AES secret key, which is sent between the tag and the central server. AES is the most widely used symmetric cryptography. This paper only discusses the standard version of Rijndael with a packet length of 128 [30]. The key exchange is the common weakness of all symmetric key methods, not just AES, so we used the ECC algorithm to cryptography AES secret key for key exchange. Therefore, we propose hybrid cryptography for mutual authentication in RFID.

The ECC algorithm is asymmetric cryptography based on an algebraic structure of elliptic curves on finite fields. The ECC is applicable for key agreement, digital signatures, pseudorandom generators, and other tasks. Indirectly, they can be used for encryption by combining the key agreement with a symmetric encryption scheme. The strength of this algorithm is its small key length, which makes it more efficient in sending keys and messages. However, the disadvantage is that the cryptographic operation is slow.

Table 1 compares the key lengths of the AES, Rabin, ECC, and RSA algorithms based on the computational effort to break them. As we can see, the AES algorithm provides the same level of security with a much smaller key length than ECC, Rabin, and RSA. We, in this paper, used a version of the AES and ECC algorithm with 128-bit keys and 233-bit keys to perform the simulation, respectively.

**Table 1 Comparison of the key size of the encryption algorithms in bits [31].**

| AES | ECC | Rabin | RSA |
|---|---|---|---|
| 56 | 112-159 | 512 | 512 |
| 80 | 160-223 | 1024 | 1024 |
| 112 | 224-255 | 2048 | 2048 |
| 128 | 256-383 | 3072 | 3072 |
| 192 | 384-511 | 7680 | 7680 |
| 256 | 512+ | 15360 | 15360 |

In our approach, First, the data is encrypted using the AES algorithm by a 128-bit secret key ($k$), then the secret key is encrypted using the ECC algorithm by the public key received from the public key management center. Finally, the encrypted text and secret key are sent to the server. The data encryption process is performed by the AES symmetric algorithm, which is faster and consumes less energy than the asymmetric algorithms. On the other hand, the AES secret key is encrypted by the ECC and sent to the server, which is more secure than symmetric methods, thus achieving speed in authentication and saving energy. As a result, we have increased the security of the secret key in symmetric encryption.

In our method, we use a mutual authentication protocol in which the tag and the central server authenticate each other. This method allows the tags to be read at high speeds. To better understand the protocol, in this research, the process of moving cars and authentication to pay for tolls can now be considered gates that drivers do not need to pay for tolls manually. Instead, they mount tags on the machine and automatically reduce the cost of their account as they pass the machine. In this research, for better understanding, Traffic volume is divided into three categories based on the number of vehicles: light, medium, and heavy.

In light traffic mode, try to use very light encryption to have short time overhead and fast authentication. In the medium mode, more robust encryption is used than before because there is little traffic in this mode, and there is no need for fast authentication. In heavy mode, because the machines are passing through the gate at a very low speed, therefore, the authentication can be done here at a lower speed, and the encoder used here can be much more robust and more secure than previously used. The protocol procedure used here is as in Figure 1.

The procedure for the proposed protocol is as follows:
1. The reader sends a 22-bit $Query$ command to the tags, and the tag that gains the slot sends its generated 16-bit RN16 to the reader to reserve the channel.
2. The reader receives $RN16$, generates a random $Cr$ value, and sends it to the tag with 18-bit $ACK$.
3. The tag encrypts its $ID$ and $Cr$ using the AES algorithm. The tag also generates a random value of $Ct$ and encrypts it along with the AES secret key ($k$) using the ECC algorithm with the public key received from the public key management center.
4. The tag sends the generated $R1$ and $R2$ messages to the reader.
5. The reader estimates the sleep time of the tag ($Time$) then sends it along with $R1$ and $R2$ to the central server.
6. The central server receives $R1$ and $R2$, first decrypting $R2$ with its private key. It then decrypts $R1$ with the key obtained from $R2$. With the value of ID ‖ Cr ‖ Ct, the ID related to the tag is found through a simple search in the database, and thus tag is authenticated.
7. The server encrypts $Cr$ and $Ct$ along with $Time$ using its secret key then sends it to the reader in order to send it for the tag.
8. After receiving this message, The tag decrypts it and check if $Cr$ and $Ct$ received was equal with its own; if true, then the tag considered that the server is valid and goes to sleep for a period of $Time$.

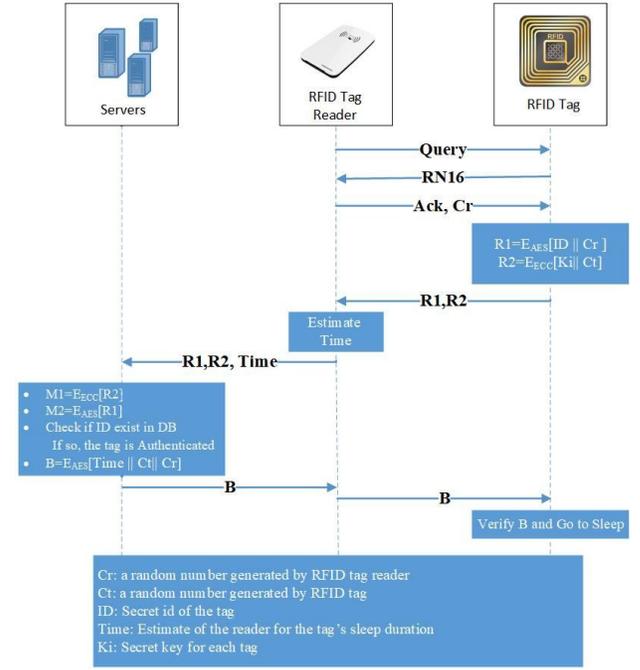

**Figure 1 Proposed mutual authentication protocol structure based on public-key cryptography**

The following assumptions have been considered for the proposed protocol:
1. The communication channel between the reader and the central server is considered secure. Therefore, this protocol addresses only threats related to the relationship between tag and reader.
2. The tags will go to sleep after the mutual authentication process is completed to avoid participating in subsequent cycles. This method has two advantages. First, tags that have been successfully authenticated do not participate in subsequent reading cycles, which reduces the likelihood of collisions for other tags. Second, this method saves battery power on the tag. Of course, the time to fall asleep is fixed, and the nodes will wake up after a certain period and search for the environment for the tag and re-enter sleep if no tag is received.
3. After receiving the encrypted message from the tag, the reader estimates the time that tag can go to sleep. The reader sends it to the central server along with the $R1$ and $R2$ messages and the values of $Cr$ and $Ct$. This estimate can be obtained based on the traffic model or the approximate distance of the car to the next reader tag. Here for the sake of simplicity, we have assumed that the tag will go to sleep as long as it is within the tag range after the authentication is complete.

### 4.2. Sleep algorithm for tags and readers

After the authentication process of a tag is completed, if the tag is still in the area of the reader, Then, the channel will again be reserved to communicate with the reader after receiving the query from the reader again. This method is especially true for moving objects with very low mobility. Like heavy traffic where tags are longer in the tag range, it manifests itself most. This scheme will reduce the chance of reserving the channel and communicating with the reader for other tags that have not yet been authenticated. Therefore, when a tag is authenticated by the central server and receives the last message from the reader, it falls into sleep mode to avoid participating in subsequent cycles.

On the other hand, we know that active RFID tags contain an internal battery for calculating and communicating with the reader. Battery life is limited to approximately 3 to 8 years. Using the sleep tags method not only improves system performance but also increases the battery life of the tags. Moreover, on the other hand, it is also useful for objects in the IoT that have limited energy, and It will extend the life of these objects. Sleep readers also improve their battery life under certain conditions. In general, the reader must sleep at a time when there is no tag within the range of the reader to save energy. Therefore, the reader should come out of the sleep in a fixed time interval to search the environment for the tags and to go to sleep if the tag is not detected.

### 5. Security analysis of the proposed protocol

The main purpose of the authentication protocol presented in this research is to reduce the complexity of computing on a central server. This means that the central server does not need to decrypt the received message for all identifiers in the database. Since there is little energy in the Internet of Things, the goal is to reduce computational overhead and thus reduce energy consumption. In this section, we analyze the security level of the proposed protocol based on informal and formal methods. Based on the informal analysis, the resistance of the proposed protocol will be evaluated against some well-known RFID attacks. Also, For the formal analysis, we use the BAN logic to evaluate the robustness of the proposed protocol.

#### 5.1. Informal analysis

In this section, we informally prove the resistance of the improved protocol against several possible attacks on the use of RFID.

#### 5.1.1 Denial of Service (DoS)

Denial of Service (DoS) is one of the severe attacks that degrade the performance of a network by disconnecting the host, bandwidth depletion, and resource depletion. This type of attack is tough to prevent because the behavior of the whole network must be analyzed. Distributed denial of service (DDoS) is a kind of DoS attack wherever multiple intermediate systems that concerned are often affected with a Trojan are used for targeting one system inflicting a DoS attack. In this DDoS attack, the network traffic comes from many various attacks [32]. On RFID, given that the mutual authentication was done. The attack of DoS and DDoS has less effect on the network.

#### 5.1.2 Impersonation

An unauthorized reader cannot access important information of tags because the secret key encrypts it. Also, this secret key is encrypted using the public key of the server so that the server only can decrypt the secret key by its private key. Likewise, no unauthorized tag can identify itself as an authorized entity in the system. Because tag IDs are only available to the central server and the tags themselves, when the server decrypts the encrypted message of the tag and finds the tag ID in the database, it ensures that the tag is allowed in the system because the unique ID of the tag is owned by itself and the central server.

#### 5.1.3 Unauthorized reading of the tag

An unauthorized reader can not be able to access the information that the tag sends to the reader, in this protocol, except for the random challenges generated by the tag and the reader. There is not any information freely published on wireless media. The tag encrypts the confidential ID on the tag using the secret key, and Also, this secret key is encrypted using the public key of the server. Therefore the central server can only open the message. Also, the attacker is not able to carry out any attacks on the system with random challenges.

#### 5.1.4 Change the tag content

No invalid reader can modify tag information by sending a signal to it Because the database signs the information that the tag receives from authorized reader tags. When the tag decrypts the message of the database, it ensures that the central server sends the message and that no changes are made. It is thus able to detect the change in the content of the tag quickly.

#### 5.1.5 Tag tracking

These attacks are one of the most controversial challenges in applying RFID technology to various applications and large scales. In this protocol, no two messages that are sent by the tag are interdependent. Each message that is encrypted by the tag contains a tag identifier and two random values that are different in each message. So, there is no way that an unauthorized reader could link messages sent to a tag and thus be able to track the tag. However, with this feature, the central server can authenticate the tag at a time $O(1)$. The strength of this protocol is that it is immune to attacks and that the computational complexity of the central server is constant.

#### 5.1.6 Resend attacks

The challenge and response phase of this protocol prevents resend attacks. For example, suppose an unauthorized tag intercepts a message from an authorized tag and wishes to send it to the reader at a later time. In the challenge and response phase, the reader generates a random number ($Cr$) and sends it to the tag as a challenge. If the tag sends the heard message to the reader, the reader sends it to the central server along with $Cr$. After decrypting the message, the server discovers the incompatibility, and as a result, the tag will not be authenticated.

### 5.1.7 Replay attacks

A replay attack is an attack in which the adversary records a communication session and replays the entire session, or some portion of the session, at a later point in time. The replayed message(s) may be sent to the same verifier as the one that participated in the original session, or to a different verifier. To counter this possibility, both the sender and receiver should establish a completely random session key, which is a type of code that is only valid for one transaction and cannot be used again. An attacker who obtains the tag answer to some query or reader answer cannot use this answer into another session because $Cr$ and $Ct$ are updated after each query. Hence, it can be said that the proposed protocol is robust against the replay attack.

### 5.2. Formal analysis based on BAN logic

In this section, we analyze the improved protocol by BAN logic method [33] [34]. BAN logic is an important tool for evaluating protocols. The used notations for this evaluation are presented in Table 2.

BAN logic evaluates protocol in four steps: Idealizing the protocol, Initiative premises, Establishment of security goals and Protocol Analysis. The rules that are used in the evaluation are as follows:

The message-meaning rule (R1)
$$\frac{P|\equiv P \overset{K}{\leftrightarrow} Q,\ P \triangleleft \{X\}_K}{P|\equiv Q|\sim X} \tag{1}$$

The nonce-verification rule (R2)
$$\frac{P|\equiv \#(X),\ P|\equiv Q|\sim X}{P|\equiv Q|\equiv X} \tag{2}$$

The jurisdiction rule (R3)
$$\frac{P|\equiv Q \Rightarrow X,\ P|\equiv Q|\equiv X}{P|\equiv X} \tag{3}$$

The freshness rule (R4)
$$\frac{P|\equiv \#(X)}{P|\equiv \#(X,Y)} \tag{4}$$

**Table 2 The notations used in the proposed protocol**

| | Notations |
|---|---|
| $P|\equiv X$ | The principal P can act if the formula X is true. |
| $\#X$ | Formula X is new and was not previously sent |
| $P \triangleleft X$ | X sent a message to P |
| $P|\sim X$ | P once said Xs |
| $P \Rightarrow X$ | P has jurisdiction on statement X |
| $\{X\}_K$ | X is encrypted using key K |
| $(A \overset{K}{\leftrightarrow} B)$ | A and B communicate using K as the shared key |

Since in this authentication process, the main entities are the server and the tag, we idealize entities as S and T. Therefore messages Message exchanged between them are as follows:

Message 1: $S \rightarrow T$: $Query$
Message 2: $T \rightarrow S$: $RN16$
Message 3: $S \rightarrow T$: $Ack, Cr$
Message 4: $T \rightarrow S$: $R1, R2$
Message 5: $S \rightarrow T$: $B$

**Step 1: Idealizing the protocol**

This step aims to convert the messages of the proposed protocol to the favorable form of BAN logic. In this step, all unencrypted messages are omitted from the protocol messages. We have an ideal form as follows:

Message 4: $T \rightarrow S$: $E\left(\left(S \overset{K_{ST}}{\leftrightarrow} T\right), Cr, ID\right), \{K_{ST}, Ct\}_{K_T}$
Message 5: $S \rightarrow T$: $E\left(S \overset{K_{ST}}{\leftrightarrow} T\right)$

**Step 2: Initiative premises**

In this step, we consider the initial assumptions of the proposed protocol concisely:

$$\begin{array}{ll} A1: S|\equiv S \overset{K_{ST}}{\leftrightarrow} T & A4: S|\equiv \#(T) \\ A2: T|\equiv S \overset{K_{ST}}{\leftrightarrow} T & A5: T|\equiv \#(Ct) \\ A3: S|\equiv \#(Cr) & A6: T|\equiv ID \end{array} \tag{5}$$

**Step 3: Establishment of security goals**

The security goals of the proposed protocol are defined as follows:

$$S|\equiv T|\sim\#\left(S \overset{K_{ST}}{\leftrightarrow} T\right),\quad T|\equiv S|\sim\#\left(S \overset{K_{ST}}{\leftrightarrow} T\right) \tag{6}$$

**Step 4: Protocol Analysis**

In this step, by applying logical rules to the initial premises and the idealized messages, we analyze the security level of the proposed protocol as follows:

Based on message 4:
$$S \triangleleft E\left(\left(S \overset{K_{ST}}{\leftrightarrow} T\right), Cr, ID\right) \tag{7}$$

Using the assumption A1 and The message-meaning rule (R1), we can conclude:
$$S|\equiv T|\sim E\left(\left(S \overset{K_{ST}}{\leftrightarrow} T\right), Cr, ID\right) \tag{8}$$

Then, based on assumptions and applying the hash rule:
$$\frac{P|\equiv Q|\sim E(X_1,X_2,...,X_n), P \triangleleft X_1, P \triangleleft X_2,..., P \triangleleft X_n}{P|\equiv Q|\sim (X_1,X_2,...,X_n)} \tag{9}$$

We can conclude that:
$$S|\equiv T|\sim\left(\left(S \overset{K_{ST}}{\leftrightarrow} T\right), Cr, ID\right) \tag{10}$$

Using the rule:
$$\frac{P|\equiv Q|\sim(X,Y)}{P|\equiv Q|\sim X} \tag{11}$$

We can deduce that:
$$S|\equiv T|\sim\left(S \overset{K_{ST}}{\leftrightarrow} T\right) \tag{12}$$

Furthermore, from the assumption A3 and The freshness rule (R4), we can deduce that:
$$S|\equiv \#\left(\left(S \overset{K_{ST}}{\leftrightarrow} T\right), Cr, ID\right) \tag{13}$$

Applying equation 10 and 13:
$$S|\equiv T|\sim\#\left(S \overset{K_{ST}}{\leftrightarrow} T\right) \tag{14}$$

As a result, the first goal has been proved; then we will prove the second goal as follow:

Based on message 5:

$$T \triangleleft E\left(S \overset{K_{ST}}{\leftrightarrow} Time\right) \quad (15)$$

Using the assumption A2 and The message-meaning rule (R1), we can conclude:

$$T|\equiv S|\sim E\left(S \overset{K_{ST}}{\leftrightarrow} T, Time\right) \quad (16)$$

Then, based on assumptions and applying the hash rule, we can conclude that:

$$T|\equiv S|\sim \left(S \overset{K_{ST}}{\leftrightarrow} T, Time\right) \quad (17)$$

Furthermore, from the assumption A5 and The freshness rule (R4), we can deduce that:

$$T|\equiv \#\left(S \overset{K_{ST}}{\leftrightarrow} T, Time\right) \quad (18)$$

Applying equation 10 and 13:

$$T|\equiv S|\sim \#\left(S \overset{K_{ST}}{\leftrightarrow} T, Time\right) \quad (19)$$

Using equation 11, we can conclude that:

$$T|\equiv S|\sim \#\left(S \overset{K_{ST}}{\leftrightarrow} T\right) \quad (20)$$

As shown above, the second goal has been proved. Therefore, we claim that the proposed protocol has high security.

## 6. Security and performance comparison

In this section, we examine the simulation scenario and describe the physical parameters intended for the simulation. Then we will present the numerical results of the simulation. Implementation is performed in the Contiki operating system. Contiki runs on types of hardware devices that are severely limited in terms of memory, power, processing power, and communication bandwidth. There is a network simulator called Cooja in every Contiki system which simulates networks of nodes in that system. Contiki is designed for small-scale systems. It has merely a few kilobytes of memory available [35] [36].

### 6.1. Simulation scenario

To investigate the protocol presented in this study, we designed a simulation scenario to observe the effect of using different algorithms on system performance. For this, we consider vehicles crossing a broad road with multiple lanes. Moreover, we assume that in the middle of the road, a reader is mounted at the height of 5 meters above the ground to read the tags of the vehicles passing by and identify that vehicle.

Since vehicles are traveling at different speeds, The limitations of these tags make it difficult for the reader to detect a large number of cars passing at high speeds. In addition to the high speed of the vehicles makes the tag low readable, using security methods also puts much overhead on the system. The simulations showed that passive tags do not perform well in this application. Because they do not have an internal battery, and these tags can never initiate the relationship between themselves and the reader. Therefore, all of the simulations in this study have been done by concerning active tags only.

The simulation performed on three models of heavy, medium, and light mobility to get a reasonable estimate of the actual traffic conditions on the roads. Because as we know, on urban and suburban roads, the volume of vehicles and the speed of their mobility vary at different hours of the day and night. So we have tried to model different traffic conditions during the day and night with these three hypothetical mobility models. The heavy traffic model is when the density of passing vehicles is high, and its speed is low. Therefore, the number of tags that are in the reader range at any given moment is too much. Moreover, each tag spends more time in the reader range. In the medium traffic model, the density of vehicles is lower while their speed is higher than heavy traffic. In this type of traffic model, vehicles are moving smoothly. In the simulations, the speed of the vehicles in this model is assumed to range from 40 to 50 km/h, indicating a normal traffic flow. The light traffic model represents the worst case. The density of vehicles in this model is lower than other models, and their speed is high. In this model, the tags are less time consuming within the reader range. Moreover, their chances of being read will be reduced. Therefore, we expect that in heavy traffic, a higher percentage of tags will be identified by the reader. Table 3 shows the characteristics of each model. This table shows the distance between the cars according to their speed, based on the time interval to reach the next car.

Table 3 Features three hypothetical traffic models

| Traffic model | Speed(m/s) | Distance between vehicles (s) | Number of lines |
|---|---|---|---|
| Light | 22-42 | 0-10 | 5 |
| Medium | 11-14 | 0.25-0.5 | 5 |
| Heavy | 1.5 | 3 | 6 |

The physical parameters have been simulated after careful examination of the structure of the tags and readers as well as the diffusion channel conditions in RFID systems in the application of urban traffic management. Table 4 shows these parameters.

Table 4 Physical parameters of the system

| parameters | Tag | Reader |
|---|---|---|
| Power | 40 mW | 10 W |
| Output power sent | -10 dBm | 0 dBm |
| Receptor sensitivity | -70 dBm | -82 dBm |
| Antenna height | 1 m | 5 m |
| Antenna type | Omnidirectional | |
| Frequency | UHF (900 MHz) | |
| diffusion Model | Small-Scale Shadowing n=4, o=2 | |
| The radius of coverage range | 10 m | |

### 6.2. The ratio of reads to tags

Figure 2 demonstrates a comparison of the percentages of tags identified by the reader in the medium traffic model for the proposed protocol and [26]. In this figure, the horizontal axis shows the packet delay between the reader and the central server in milliseconds. Moreover, the vertical axis represents the proportion of tags that have been successfully authenticated. The figure has been shown for bandwidth 256 kbps between the tag and the reader.

The figure shows that Due to the high speed of the proposed authentication protocol than [26], the fraction of the tags read for our scheme is high. The reason is that each key-ring contains $t$ blocks. Also, it can be concluded that the more the bandwidth of the wireless channel between the tag and the reader, The higher percent of tags will be identified when passing through the reader. The reason is that the more bandwidth, the delay to read a tag comes down. Moreover, as a result, other tags in the range will have more opportunities to connect with the reader.

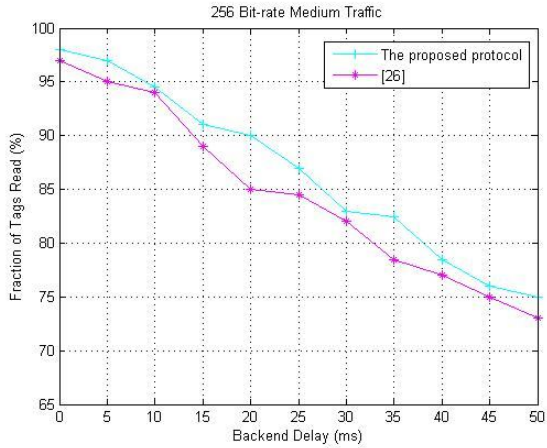

**Figure 2 Ratio of tag reads in a medium traffic model with bit-rate 256 kbps between the proposed protocol and [26] .**

### 6.3. Energy consumption of tags

In our scheme, we have used the idea of sleeping tags after successful authentication in the simulation of the system. In this case, a tag goes to sleep when its authentication is complete; it not only saves energy but also increases the chances of other tags being read in the environment. In table 5, we observe the average percentage of activity of a tag over the range of the reader for different delays with considering bandwidth 1 Mbps for the proposed protocol and [26]. Results show that in our approach, tags are only about 18 to 26 percent of their time in the reader area and spend most of their time in sleep.

**Table 5 The effect of sleeping on tags in different traffic models assuming 1Mbps bandwidth**

| Traffic model | Delay reading a tag (ms) | | Time in range (seconds) | | The ratio of awakening time to total time in the coverage area for various delays | | | |
|---|---|---|---|---|---|---|---|---|
| | | | | | 0 ms | | 25 ms | |
| | [26] | A | [26] | A | [26] | A | [26] | A |
| Light | 1.7 | 1.3 | 0.718 | 0.718 | 20.27% | 18.31% | 21.72% | 18.16% |
| Medium | 1.7 | 1.3 | 1.612 | 1.612 | 20.18% | 20.92% | 21.03% | 19.64% |
| Heavy | 1.7 | 1.3 | 13.333 | 13.333 | 25.67% | 23.45% | 25.96% | 22.39% |

A: The proposed protocol

In the proposed protocol, successful read latency for a tag with considering 1 Mbps bandwidth and using the hybrid cryptography algorithm is approximately 1.3 milliseconds regardless of the latency between reader and server. This delay is equal to the time interval between receiving a request from the reader to receiving the last message from the reader in the same cycle and completing mutual authentication. Also, in the simulations, the average presence of a tag in the range of the reader for heavy, medium, and light traffic models were reported to be 13.334, 1.612, and 0.718 seconds, respectively.

### 6.4. The comparison with other protocols

In this section, we will analyze the simulation results of the authentication protocol implemented by the hybrid cryptography algorithm with a 128-bit key length for the AES algorithm. It should be noted that with this key length, the Rabin encryption algorithm will not provide relatively high security for the system. To get an overview of the conditions for deciding on system security issues, and based on the type of application and other considerations, the most appropriate algorithm and parameters can be selected for the system.

In Table 6, we compare the proposed protocol to other protocols based on security properties. In Section 5, we proved that the proposed protocol is robust against attacks in table 6.

**Table 6 The security comparison of the proposed protocol to other protocols**

| Protocols | A1 | A2 | A3 | A4 | A5 |
|---|---|---|---|---|---|
| Yuan and Liu [37] | ✓ | ✓ | ✓ | × | ✓ |
| Hsi et al. [38] | ✓ | ✓ | ✓ | × | ✓ |
| Rahman et al. [39] | ✓ | ✓ | × | ✓ | ✓ |
| Vijaykumar et al. [40] | ✓ | ✓ | × | × | ✓ |
| Sundaresan et al. [41] | × | ✓ | ✓ | × | ✓ |
| Chen et al. [42] | ✓ | ✓ | × | × | ✓ |
| Proposed protocol | ✓ | ✓ | ✓ | ✓ | ✓ |

A1 Denial of Service (DoS) attack, A2 Impersonation attack, A3 Replay attack, A4 Tag tracking attack, A5 Resend attacks, ✓ Resistant × Non-resistant

## 7. Conclusion

The main idea of designing the proposed protocol is to reduce the computational complexity of the central server. In traffic management applications on RFID, there may be a large number of requests for authentication from different readers at any one time to the central server, which is a significant challenge for these systems. On the other hand, resource constraints in these systems make it challenging to use powerful public-key cryptographic algorithms. Mobility in the system and the need to quickly identify tags also add time constraints to existing challenges. Security against tag tracking attacks is an essential feature of the proposed protocol because this protocol is designed to reduce the server-side computations from $O(n)$ to $O(1)$ in addition to the privacy of the tags. Because in the proposed protocol, every two messages sent to the server are independent of each other, and each encrypted message contains a tag identifier and two random values that differ in each message. As a result, the central server can authenticate the tag at a constant time $O(1)$.

The simulation results show that with the low bit-rate Rabin-512 encryption algorithm, we are also able to detect a high percentage of tags in the environment. At a rate of 128 kb/s, we can detect more than 90% of tags for delays of up to 10 milliseconds between the reader and server. Regarding the energy consumption of the tag, it was shown that using the idea of sleeping on the tags can save up to 75% to 80% on the energy consumption of the tags. Also, sleeping the reader during idle time can significantly extend its battery life by maintaining proper system performance.